\documentclass[a4paper,fleqn]{cas-dc}

\usepackage[numbers]{natbib}

\usepackage{xcolor}
\usepackage{amsmath}
\usepackage{amsfonts}
\usepackage{graphicx}
\usepackage{xparse}
\usepackage{mathtools}
\usepackage{stmaryrd}
\usepackage[ruled,vlined]{algorithm2e}
\usepackage{amsthm} 
\usepackage{caption}

\usepackage{amsmath}
\usepackage{amssymb}
\usepackage{algorithmic}

\hypersetup{
    colorlinks=true,
    linkcolor=black,
    urlcolor=blue,
    pdftitle={},
    pdfpagemode=FullScreen,
    }
\usepackage{diagbox}

\def\tsc#1{\csdef{#1}{\textsc{\lowercase{#1}}\xspace}}
\tsc{WGM}
\tsc{QE}
\tsc{EP}
\tsc{PMS}
\tsc{BEC}
\tsc{DE}

\begin{document}
\let\WriteBookmarks\relax
\def\floatpagepagefraction{1}
\def\textpagefraction{.001}
\shorttitle{Unveiling Free Robustness Margins}
\shortauthors{Y.A Si et~al.}

\newtheorem{theorem}{Theorem}[section]
\newtheorem{conjecture}[theorem]{Conjecture}
\newtheorem{assumption}[theorem]{Assumption}
\newtheorem{corollary}[theorem]{Corollary}
\newtheorem{proposition}[theorem]{Proposition}
\newtheorem{definition}[theorem]{Definition}
\newtheorem{lemma}[theorem]{Lemma}
\newtheorem{remark}[theorem]{Remark}
\newtheorem{problem}[theorem]{Problem}

\newcommand{\Ze}{{\mathbb Z}}
\newcommand{\R}{{\mathbb{R}}}
\newcommand{\N}{{\mathbb{N}}}
\newcommand{\ie}{{\it i.e.,}\xspace}
\newcommand{\segcc}[1]{{{\left\llbracket#1\right\rrbracket}}}
\newcommand{\Pre}{\textrm{Pre}}
\newcommand{\Inc}{\textrm{Inc}}
\newcommand{\dom}{\textrm{dom}}
\newcommand{\argmax}{\textrm{Argmax}}
\newcommand{\RM}{\textrm{RM}}
\newcommand{\cl}{\textrm{cl}}

\newcommand{\Int}{\textrm{Int}}

\title [mode = title]{Symbolic Control: Unveiling Free Robustness Margins}                      



\author[1]{Youssef Ait Si}[style=chinese]
\ead{youssef.aitsi@um6p.ma}


\affiliation[1]{organization={College of Computing, University Mohammed VI Polytechnic},
                addressline={Lot 660, Hay Moulay Rachid}, 
                city={Benguerir},
                postcode={695013}, 
                country={Morocco}}

\author[2]{Antoine Girard}[style=chinese]

\ead{antoine.girard@centralesupelec.fr}


\affiliation[2]{organization={Université Paris-Saclay, CNRS, CentraleSupélec, Laboratoire des Signaux et Systèmes},
                postcode={91190}, 
                city={Gif-sur-Yvette},
                country={France}}

\author[1]{Adnane Saoud}[style=chinese]
\ead{adnane.saoud@um6p.ma}

\begin{abstract}
This paper addresses the challenge of ensuring robustness in the presence of system perturbations for symbolic control techniques. Given a discrete-time control system that is related to its symbolic model by an alternating simulation relation. In this paper, we focus on computing the maximum robustness margin under which the symbolic model remains valid for a perturbed-version of the discrete-time control system. We first show that symbolic models are inherently equipped with a certain free robustness margins. We then provide constructive procedures to compute uniform and non-uniform (state and input dependent) robustness margins. We also show that the tightness of the robustness margin depends on the tightness of the reachability technique used to compute the symbolic model. We then explain how the computed robustness margin can be used for the sake of controller synthesis. Finally, we present two illustrative examples to demonstrate the effectiveness of our approach.

\end{abstract}




\sloppy

\maketitle

\section{Introduction}
Recent studies have shown importance of designing controllers with correctness guarantees for complex specifications, given their growing importance in cyber-physical systems \cite{tabuada2009verification}, \cite{belta2017formal}. One  approach to achieve this is Abstraction-Based Controller Design \cite{ReissigWeberRungger17}, \cite{DBLP:conf/hybrid/HsuMMS18}, \cite{saoud2019compositional} which can be done in three steps. In the first step, we discretize the state and input spaces of a non-linear continuous-space dynamical system and generate a finite state abstraction using tools from reachability analysis \cite{althoff2021set}. This abstraction simplifies the system by approximating its behavior in a finite, symbolic form. Next, we synthesize controllers ensuring the satisfaction of a given complex temporal logic specification. Finally, the discrete controller is translated into a continuous controller for the original dynamical system, preserving its correctness in the continuous domain. This three-step process allows for the rigorous design of controllers while guaranteeing both safety and performance properties. The correctness of these techniques rely on the use of the alternating simulation or feedback refinement relationship \cite{ReissigWeberRungger17}, \cite{DBLP:conf/concur/AlurHKV98}, \cite{liu2016finite}, \cite{NilssonOL17} between the original system and its abstraction.

In the context of controller synthesis under complex temporal specifications for dynamical systems subject to modeling uncertainties or disturbances, various approaches have been proposed in the literature. The authors in \cite{ehlers2014resilience} explore the robustness of a system in satisfying a given assume-guarantee contract with respect to assumption's violation. In the work of \cite{tabuada2014towards,rungger2015notion,apaza2024synthesis}, the robustness of a given system ensures that perturbed trajectories remain close to disturbance-free trajectories with respect to a given metric. In \cite{Adnane2024}, the authors propose a new resilience metric for  autonomous discrete-time systems under temporal logic specifications, where this metric can be computed in terms of robust optimization tools. As for abstraction-based approaches, these can be modified to account for modeling uncertainties and disturbances, by considering an upper bound on the uncertainty, and by over-approximating the worst-case scenario when computing the reachable sets \cite{ReissigWeberRungger17}, \cite{DBLP:conf/hybrid/HsuMMS18}. In this context, the authors in \cite{liu2016finite} introduce finite abstractions with robustness margins to synthesize controllers for dynamical systems from temporal logic specifications, addressing issues like intersample behavior, imperfect state measurements, and unmodeled dynamics. To improve upon these potentially conservative approaches, the authors in \cite{bai2019incremental,bai2020accurate} propose an adaptation procedure for system abstractions based on state and input dependent perturbation bounds. Additionally, they introduce an incremental algorithm to locally refine the abstractions in response to changes in the disturbance model, targeting only the affected regions of the state space.

While all the existing robust abstraction-based approaches proposed in the literature make it possible to deal with a predefined set of disturbances, we address in this paper the following question: given a dynamical system and its symbolic abstraction, what is the largest disturbance under which the symbolic abstraction (originally computed for the disturbance-free system) is still valid, in the sense that it conserves its behavioral relationship with the perturbed discrete-time system. This maximal admissible disturbance can be interpreted as a measure of robustness for the dynamical system. Indeed, the main intuition behind the proposed concept of robustness is based on the fact that a symbolic model represents not only an abstraction of a single dynamical system but of a set of dynamical systems. Consequently, controllers synthesized through symbolic approaches are inherently equipped with a certain free robustness margins. We first show formally that for a general nonlinear discrete-time control system, the constructed symbolic abstraction is equipped with a free-robustness margin. We then provide a constructive procedure to compute an upper bound on the admissible robustness margin. We also show that the tightness of the computed robustness margin depends on the tightness of the over-approximation of the reachable set. The paper analyses two types of robustness margins, non-uniform robustness margins that are state and input dependent, and uniform robustness margins. Finally, we explain how the computed robustness margin can be exploited for the sake of controller synthesis, by showing how to refine the controller from the disturbance-free symbolic abstraction to a perturbed version of the original system. Two numerical examples are proposed to show the performance of the proposed approach, a double integrator with a simple safety specification, and a nonlinear model of a mobile robot with a more complex Co-Safe LTL specification. 

This paper is organized as follows. In Section \ref{sec:2}, some
required preliminaries on transition systems and symbolic abstractions are provided. In Section \ref{sec:3}, we provide the main results of the paper by demonstrating how to compute the non-uniform and uniform robustness margins. Finally, in
Section \ref{sec:4}, two illustrative example are proposed demonstrating the validity of our approach.

\section{Preliminaries}
\label{sec:2}
\textbf{Notations:}
The symbols $ \N $, $ \N_{\geq 0} $, $ \R$, and $\R_{\geq 0},$ denote the set of positive integers, nonnegative integers, real and non-negative real numbers, respectively. Given sets $X$ and $Y$, we denote by $f: X \rightarrow Y$ an ordinary map from $X$ to $Y$, whereas $f : X \rightrightarrows Y$ denotes a set valued map. The notation \( 2^{ X} \) represents the power set of \(  X \), which is the set of all possible subsets of \( X \). For a set $X$, $\cl(X)$ denotes its closure, $\Int(X)$ its interior and $\partial( X)$ its boundaries. For $x \in \mathbb{R}^n$, the infinity norm of $x$ is denoted $\lVert x\rVert$. The closed ball centered at $x \in \mathbb{R}^n$ with radius $r$ is defined by $\mathcal{B}_{r}(x) = \{y \in \mathbb{R}^n |\lVert x- y\rVert \leq r  \}$.

\subsection{Transition systems}

First, we review the notion of \textit{transition system}~\cite{tabuada2009verification}. This notion allows us to describe the concrete dynamical system and its symbolic abstraction in a unified framework.
\begin{definition}\label{definition6}
A transition system is a tuple 
$S=(X,X^0,U,\Delta)$, where $ X $ is the set of states, $X^0 \subseteq X$ is the set of initial states, $U$ is the set of inputs and $ \Delta \subseteq X\times U\times X $ is the transition relation. When $ (x,u,x')\in\Delta$, we use the alternative representation  $ x'\in \Delta(x,u),$ where state $ x' $ is called a $ u $-successor (or simply successor) of state $ x $ under input $ u\in U $. 
\end{definition}


Given $x\in X$, the set of enabled (admissible) inputs for $x$ is denoted by $U^a(x)$ and defined as $U^a(x)=\{u\in U \mid \Delta(x,u)\neq \emptyset\}$. The transition system $S$ is said to be non-blocking if for all $x \in X$, $U^a(x)\neq \emptyset$. 

    A behavior of the transition system $S=(X,X^0,U,\Delta)$ is a sequence of states and inputs \\ $ \sigma=((x_0,u_0),(x_1,u_1),(x_2,u_2)\ldots, x_n) $, where \( n \in \N \cup \{ + \infty\}\) with $
 u_i \in U^a(x_i)$ and $x_{i+1} \in \Delta(x_i,u_i)$ for all $i \in \{0,1,\ldots,n-1\}$.
    The corresponding state-trajectory of $\sigma$ is given by $\sigma_x = (x_0,x_1,x_2, \ldots, x_n)$. A behavior $ \sigma$ is said to be: 
 \begin{itemize}
    \item  $\operatorname{maximal}$, if it does not exist a behavior $\sigma'=((x_0',u_0'),(x_1',u_1'),(x_2',u_2')\ldots, x_m')$ with $m > n$ and such that for all $  i\in \{0,1,\ldots,n\}$, $x_i=x_i'$ and $u_i=u_i'$. 
    \item $\operatorname{complete}$, if \( n = + \infty\)
\end{itemize}
The set of all maximal behaviors of $S$ is called the behavior of $S$ and denoted $\mathcal{H}(S)$. The corresponding set of all maximal state trajectories of $S$ is denoted $\mathcal{H}_{x}(S)$. A specification for the transition system $S=(X,X^0,U,\Delta)$ is denoted as $\mathcal{H}_{Spec} \subseteq  \mathcal{H}_{x}(S)$ and represents a set of state trajectories for the considered system.

In the sequel, we consider the relationship for transition systems based on the notion of alternating simulation relation to relate abstractions to concrete systems. 

\begin{definition}
	\label{Def:altsimu}
	Let $S_1=(X_1,X_1^0,U_1,\Delta_1)$ and $S_2=(X_2,X_2^0,U_2,\Delta_2)$ be two transition systems such that $X_1 \subseteq X_2$ and $U_2 \subseteq U_1$. A relation $\mathcal R \subseteq X_1\times X_2$ is said to be an alternating simulation relation from $S_2$ to $S_1$, if it satisfies:
	\begin{itemize}
		\item[(i)] $\forall x_2^0\in X_2^0$, $\exists  x_1^0\in X_1^0$ such that $(x_1^0,x_2^0)\in \mathcal{R}$;
		\item[(ii)]  $\forall (x_1,x_2)\in \mathcal R$, $\forall u_2 \in U^a_2(x_2)$, we have
		$ u_2 \in U^a_1(x_1)$ and $ \forall x_1'\in  \Delta_1(x_1,u_2)$, $\exists x_2'\in  \Delta_2(x_2,u_2)$ satisfying $(x_1',x_2')\in \mathcal R$.
	\end{itemize}
\end{definition}

\medskip

We denote the existence of an \textit{alternating simulation relation} from \( S_2 \) to \( S_1 \) by \( S_2 \preccurlyeq S_1 \).  This relation signifies that the behaviors (or trajectories) of the abstract system \( S_2 \) can effectively mimic those of the original system \( S_1 \). As a result, any discrete controller synthesized for the abstraction \( S_2 \) can be systematically refined into a hybrid controller for the original system \( S_1 \). This property enables correct-by-construction controller design based on the abstraction \cite{tabuada2009verification, belta2017formal}.

\subsection{Symbolic abstractions of dynamical systems}
\label{Def:Sd}

In this paper, we consider a discrete-time control system \(\Sigma\), defined as:
\begin{equation}
\label{eqn:system}
x_{k + 1} = f(x_k, u_k, d_k)
\end{equation}
where \( x_k \in X \subseteq \mathbb{R}^n \) denotes the state, \( u_k \in U \subseteq \mathbb{R}^p \) the control input, and \( d_k \in D \subseteq \mathbb{R}^q \) the disturbance input. Furthermore, for a map \( \varepsilon: X \times U \rightarrow \mathbb{R}_{\geq 0} \), we introduce the concept of an \( \varepsilon \)-perturbed version of \(\Sigma\), denoted as \( \Sigma_{\varepsilon} \) and formally defined as follows:
\begin{equation}
x_{k+1} \in f(x_k, u_k, d_k) + \mathcal{B}_{\varepsilon(x_k, u_k)}(0)
\end{equation}
where \( \mathcal{B}_{\varepsilon(x_k, u_k)}(0) \) denotes a ball of radius \( \varepsilon(x_k, u_k) \) around the origin.

In the following, we model the dynamical system $\Sigma$ as a transition system \( S(\Sigma) = (X, X^0, U, \Delta) \). This makes it possible to have a description of the original dynamical system and its symbolic abstraction in a unified framework. The transition system $S(\Sigma)$ is defined as follows: 
\begin{itemize}
    \item \( X \subseteq \mathbb{R}^n \) represents the set of states,
    \item \( X^0 \subseteq X \) is the set of initial states,
    \item \( U \subseteq \mathbb{R}^p \) is the set of inputs,
    \item \( \Delta: X \times U \rightarrow X \) is a nondeterministic transition relation defined for $x \in X$ and $u \in U$ as \( x' \in \Delta(x, u) \) if \( x' \in f(x, u, D) \).
\end{itemize}

Similarly for $\varepsilon:X\times U \rightarrow \mathbb{R}_{\geq 0}$, the perturbed system $\Sigma_{\varepsilon}$ can be represented as a transition system $S(\Sigma_{\varepsilon})=(X,X^0,U,\Delta_{\varepsilon})$, where $x'\in \Delta_{\varepsilon}(x,u)$ if $x' \in f(x,u,D)+\mathcal{B}_{\varepsilon(x, u)}(0)$.

The symbolic model \( S_d(\Sigma) = (X_d, X_d^0, U_d, \Delta_d) \) of the system \( S(\Sigma) \) is constructed as follows:
\begin{itemize}
    \item \( X_d = \{ q_i \mid i = 0, 1, \ldots, N \} \), where $q_0=\mathbb{R}^n \setminus X$ and $\{ q_i \mid i = 1, 2, \ldots, N \}$ is a finite partition of \( X \),
    \item \( X_d^0 \subseteq X_d \), the set of initial symbolic states,
    \item \( U_d = \{ v_h \mid h = 1, \ldots, M \} \subseteq U \), is a finite set of symbolic inputs,
    \item \( \Delta_d \subseteq X_d \times U_d \times X_d \) is a transition relation defined for $q,q' \in X_d$ and $v \in U_d$ as follows: 
    \begin{itemize}
        \item For $q \in X_d \setminus \{q_0\}$,  \( q' \in \Delta_d(q, v) \) if \( \cl(\overline{f}(q, v, D))   \cap \cl(q') \neq \emptyset \), where \( \overline{f}(q, v, D) \) is an overapproximation of the reachable set \( f(q, v, D) \) i.e, $ f(q, v, D) \subseteq \overline{f}(q, v, D)$;
        \item $\Delta_d(q_0,v)=X_d$ for all $v \in U_d$.
\end{itemize}

\end{itemize}
Using this construction of the symbolic model $S_d(\Sigma)$, one can ensure the existence of an alternating simulation relation from \( S_d(\Sigma) \) to \( S(\Sigma) \) \cite{tabuada2009verification}.

Additionally, a \textit{quantizer} \( Q: X \rightarrow X_d \) assigns each state \( x \in X \) to its corresponding symbolic state \( q \in X_d \), ensuring \( x \in q \). We also define the inverse map of the quantizer $Q^{-1}:X_d \rightrightarrows X$ defined for $q \in X_d$ as $Q^{-1}(q)=\bigcup\limits_{x\in q}\{x\}$. Similarly, for a set $A \subseteq X_d$, $ Q^{-1}(A)=\bigcup\limits_{q \in A}Q^{-1}(q)$. The inverse map of the quantizer can also be generalized to the specification set as follows, for an abstract specification $\mathcal{H}_{Spec}\subseteq \mathcal{H}_{x}(S_d)$, the corresponding concrete specification is $Q^{-1}(\mathcal{H}_{Spec}) \subseteq \mathcal{H}_{x}(S)$ and defined as $Q^{-1}(\mathcal{H}_{Spec})=\{\sigma_x=x_0,x_1,\ldots,x_n \in \mathcal{H}_{x}(S) \mid \sigma_{d,x}= Q(x_0),Q(x_1),\ldots,Q(x_n) \in  \mathcal{H}_{Spec}\}$. 
Now given the discrete-time control system $\Sigma$ and its symbolic model $S_d(\Sigma)$ constructed above, we define the map $\eta:X_d \times U_d \times 2^{X_d} \rightarrow \mathbb{R}_{\geq 0}$ representing the maximum robustness margins of $(q, v) \in X_d \times U_d$  to reach the set $A \subseteq X_d$:

\small
\begin{equation}
\label{Def:eta_a}
\eta(q, v, A)= 
\begin{cases}
\begin{aligned}[t]
 &  \sup \{ \varepsilon \geq 0 \mid  \cl(\overline{f}(q,v,D)) + \mathcal{B}_{\varepsilon}(0)\\ & \hspace{3.4cm} \subseteq {\Int(Q^{-1}(A)}) \}  \\
&  \qquad  \qquad \text { if } \cl(\overline{f}(q,v,D)) \subseteq {\Int(Q^{-1}(A))}  \\
& 0  \qquad \qquad \text {Otherwise}  
\end{aligned}
\end{cases}
\end{equation}

\normalsize

Intuitively, \(\eta(q, v, A)\) measures how much we can expand the over-approximation of the reachable set $\overline{f}(q,v,D)$ while ensuring the resulting set remains entirely within the interior of the set \(Q^{-1}(A) \subseteq X\).

\section{Robustness margins for symbolic abstractions}
\label{sec:3}

In this section, we provide our main result, by analyzing the robustness margin of the constructed symbolic models. 

\subsection{Non-uniform robustness margins}

Before providing the main result of this section, we first have the following auxiliary property.

\begin{lemma}
\label{lem:1}
    Consider the discrete-time control system $\Sigma$ in (\ref{eqn:system}) and its associated symbolic model $S_d(\Sigma)=(X_d, X_d^0, U_d, \Delta_d)$ constructed in section \ref{Def:Sd}. Then, for all $(q,u) \in X_d \times U_d$, we have $\cl(\overline{f}(q,u,D)) \subseteq \Int(Q^{-1}(\Delta_d(q,u)))$.
\end{lemma}
\begin{proof}
Consider $q\in X_d$ and $u \in U_d$, we have from the construction of the transition relation $\Delta_d$ that $\cl(\overline{f}(q,u,D)) \subseteq \cl(Q^{-1}(\Delta_d(q,u)))$. Now consider $x \in \cl(\overline{f}(q,u,D))$ and assume that $x \in \partial (Q^{-1}(\Delta_d(q,u)))$, which implies the existence of 
      \begin{equation}
    \label{eqn:2}
        q'\in X_d \setminus Q^{-1}(\Delta_d(q, u)),
    \end{equation}
with $x \in \cl(q')$. The latter implies that $\cl(\overline{f}(q,u,D)) \cap \cl(q') \neq \emptyset$. Hence, one gets $q' \in \Delta_d(q,u)$, which contradicts (\ref{eqn:2}). Hence, $\cl(\overline{f}(q,u,D)) \cap \partial(Q^{-1}(\Delta_d(q,u)))=\emptyset$ and $\cl(\overline{f}(q,u,D)) \subseteq \Int(Q^{-1}(\Delta_d(q,u)))$. 
\end{proof}
Building on Lemma \ref{lem:1} above, we can now state the main result of this paper, which allows us to characterize the admissible robustness margin.
\begin{theorem}
    \label{thrm:main}
    Consider the discrete-time control system $\Sigma$ in (\ref{eqn:system}) and its associated symbolic model $S_d(\Sigma)=(X_d, X_d^0, U_d, \Delta_d)$ constructed in section \ref{Def:Sd}.
    Consider the relation $\mathcal{R} \subseteq X\times X_d$ defined by 
    \begin{equation}
    \label{eqn:relation11}
        (x, q) \in \mathcal{R} \text{ if and only if } x \in q.
        \end{equation}
    Consider the map $\varepsilon:X \times U_d \rightarrow \mathbb{R}_{\geq 0}$, defined for $(x,u) \in X \times U_d$ as $\varepsilon(x,u)=\eta(Q(x),u,A)$, with $A=\Delta_d(Q(x),u)$. The following properties hold:
    \begin{itemize}
        \item  The map $\varepsilon:X \times U_d \rightarrow \mathbb{R}_{\geq 0}$ satisfies $\varepsilon(x,u) >0$ for all $(x,u) \in X\times U_d$;
        \item For any map $\mu:X \times U_d \rightarrow \mathbb{R}_{\geq 0}$ satisfying 
        \begin{equation}
        \label{eqn:robu1}
             \mu(x,u) < \varepsilon(x,u) \text{ for all } (x,u) \in X \times U_d,
        \end{equation}
        The relation $\mathcal{R}$ in (\ref{eqn:relation11}) is an alternating simulation relation from $S_d(\Sigma)$ to  $S(\Sigma_{\mu})$;
        \item For $\delta \geq 0$, assume that $\overline{f}$ is a $\delta$-overapproximation\footnote{The overapproximation  $\overline{f}:X\times U \times D \rightrightarrows X$  is a $\delta$-overapproximation of the map $f:X\times U\times D \rightarrow X$ if for all $x \in X$ and for all $u \in U$, $\overline{f}(x,u,D) \subseteq f(x,u,D) + \mathcal{B}_{\delta}(0)$.} of the map $f$ describing the dynamics of the system $\Sigma$ and that the set $\overline{f}(q,v,D)$ is closed for all $(q,v) \in X_d \times U_d$. Consider a map $\overline{\mu}:X \times U_d \rightarrow \mathbb{R}_{\geq 0}$ satisfying the following:

        \begin{equation}
        \label{eqn:robu2}
        \exists q\in X_d, \exists u \in U_d, \text{s.t.}, \forall x\in q, \quad \overline{\mu}(x,u) > \varepsilon(x,u)+\delta.
        \end{equation}
        Then, the relation $\mathcal{R}$ in (\ref{eqn:relation11}) is not an alternating simulation relation from  $S_d(\Sigma)$ to  $S(\Sigma_{\overline{\mu}})$.      
        \end{itemize}
\end{theorem}
\begin{proof} 
We provide a proof for each item separately. Consider $(x,u) \in X \times U_d$ and let $q=Q(x) \in X_d$. We have from Lemma \ref{lem:1} that $\cl(\overline{f}(q,u,D)) \subseteq \Int(Q^{-1}(A))$, which implies from the fact that the set $\Int(Q^{-1}(A))$ is open 
the existence of  an $ \alpha>0$ such that $\cl(\overline{f}(q,u,D))+\mathcal{B}_{\alpha}(0) \subseteq \Int(Q^{-1}(A))$, which in turn implies from the definition of $\eta(q,u,A)$ in (\ref{Def:eta_a}) that $\varepsilon(x,u)=\eta(Q(x),u,A)>0$.

To show the behavioral relationship between $S_d(\Sigma)$ and $S(\Sigma_{\mu})$, consider a map $\mu:X \times U_d \rightarrow \mathbb{R}_{\geq 0}$ satisfying $\mu(x,u) < \varepsilon(x,u)$ for all $(x,u) \in X \times U_d$.  Let us now prove the first condition in Definition \ref{Def:altsimu}.
Consider a discrete state $q \in X_d^0$. Since $X_d^0 \subseteq X^0$, it follows that $q \subseteq X^0$.
By the non emptiness of $q$ there exists at least one element $ x \in q \subseteq X^0$. Thus, there exits $x \in X^0$ such that $x \in q$.

    For the second condition, let us consider $(x, q) \in \mathcal{R}$ and $v \in U^a(q)$. 
By the definition of the enabled inputs, $q$ has at least one successor under the input $v$.
We have $v \in U^a(x)$ and consider any $x' \in \Delta_{\mu}(x,v)$ i.e. $x' \in f(x,v,D) + \mathcal{B}_{\mu(x, v)}(0)$. We have that $x' \in f(q, v, D) + \mathcal{B}_{\mu(x,v)}(0) \subseteq \cl(\overline{f}(q,v,D)) +  \mathcal{B}_{\mu(x, v)}(0) \subseteq \cl(\overline{f}(q,v,D)) +  \mathcal{B}_{\varepsilon(x, v)}(0) $. Moreover, we have $\varepsilon(x, v) = \eta (Q(x), v,A) = \eta(q, v,A)$ with $A= \Delta_d(q,v)$. Hence, by the definition of the map $\eta$ in (\ref{Def:eta_a}), one gets $\cl(\overline{f}(q,v,D)) +  \mathcal{B}_{\varepsilon(x, v)}(0) \subseteq \Int(Q^{-1}(\Delta_d(q,v)))$. Thus, there exists $ q' \in \Delta_d(q,v)$ where $x' \in q'$. This implies  $(x', q') \in \mathcal{R}$, and condition (ii) in Definition  \ref{Def:altsimu} holds, which concludes the proof for  $S_d(\Sigma)\preccurlyeq S(\Sigma_{\mu})$.

To show that $\mathcal{R}$ in (\ref{eqn:relation}) is not an alternating simulation relation from  $S_d(\Sigma)$ to $S(\Sigma_{\overline{\mu}})$, we choose to establish that the second condition in Definition  \ref{Def:altsimu} does not hold.

Consider a map $\overline{\mu}:X \times U_d \rightarrow \mathbb{R}_{\geq 0}$ and let $q\in X_q$ and $u\in U_d$ such that for all $x\in q$, $\overline{\mu}(x,u) > \varepsilon(x,u) + \delta$. By definition of the map $\eta$ in (\ref{Def:eta_a}), we have the existence of $z \in q$ such that for all $\varepsilon_0 > \varepsilon(z,u)$
\begin{equation}
\{\cl(\overline{f}(z, u, D)) +  \mathcal{B}_{\varepsilon_0}(0) \}  \cap \mathbb{R}^n \setminus Q^{-1}(\Delta_d(q, u)) \neq \emptyset.  
\end{equation}
In particular for $\varepsilon_0=\bar{\mu}(z,u)-\delta> \varepsilon(z,u)$, we have that: 
\begin{equation}
\label{eqn:relation3}
\{\cl(\overline{f}(z, u, D)) +  \mathcal{B}_{\bar{\mu}(z,u)-\delta}(0) \} \cap \mathbb{R}^n \setminus Q^{-1}(\Delta_d(q, u)) \neq \emptyset  
\end{equation}
Moreover, since $\overline{f}$ is a $\delta-$overapproximation of $f$ and the overapproximation map $\overline{f}$ is closed, we have:
\begin{align*}
\cl(\overline{f}(z,u,D)) + \mathcal{B}_{\bar{\mu}(z,u)-\delta}(0)  &= \overline{f}(z,u,D) + \mathcal{B}_{\bar{\mu}(z,u)-\delta}(0) \\  &\subseteq f(z,u,D) + \mathcal{B}_{\bar{\mu}(z,u)}(0)
\end{align*}
Combining the last inclusion with (\ref{eqn:relation3}) we deduce that $  f(z,u,D) +  \mathcal{B}_{\overline{\mu}(z, u)}(0) \cap \mathbb{R}^n \setminus Q^{-1}(\Delta_d(q, u)) \neq \emptyset$. Hence, there exists $y \in \{f(z,u,D)+\mathcal{B}_{\overline{\mu}(z,u)}(0)\}$ such that for all $q' \in \Delta_d(q,u)$, $y \notin q'$. Thus, $\mathcal{R}$ is not an alternating simulation relation from $S_d(\Sigma)$ to $S(\Sigma_{\overline{\mu}})$.

\end{proof}

\begin{figure}[!tbp]
  \centering
  \begin{minipage}[b]{0.35\textwidth}
  \begin{center}
    \includegraphics[width=\textwidth]{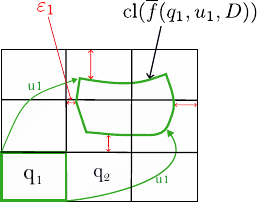}
    \end{center}
  \end{minipage}
  \hfill
  \vspace{0.2cm}
  \begin{minipage}[b]{0.4\textwidth}
  \begin{center}
    \includegraphics[width=0.9\textwidth]{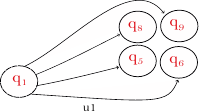}
    \end{center}
  \end{minipage}
  \caption{The first figure displays the overapproximation of the reachable set for the discrete state \( q_1 \) under the discrete input $u_1$. It also illustrates the margin $\eta(q_1,u_1,\Delta_d(q_1,u_1))=\varepsilon_1$ that can be added to the borders of this overapproximation while preserving the same set of successors. The second figure shows the equivalent automata for the transitions from the discrete state \( q_1 \) under the discrete input \( u_1 \).}
   \label{fig:robustness_margin}
\end{figure}

The result of the previous theorem shows that symbolic abstractions are inherently equipped with a certain free robustness margins. The result provides also an approach to compute the admissible robustness margins for the constructed symbolic abstraction. While the classical results in the symbolic control literature \cite{tabuada2009verification, ReissigWeberRungger17,zamani2011symbolic} have shown the existence of an alternating simulation relation between the concrete system and its symbolic abstraction, the proposed result moves one step forward by providing a collection of disturbed systems under which the computed symbolic abstraction (for the disturbance-free system) is still valid, in the sense that it conserves its behavioral relationship with the perturbed discrete-time control system. The previous result also shows that the tightness of the obtained robustness margin depends on the tightness of the overapproximation of the reachable. In particular, when the computed reachable set is exact, i.e, $\delta=0$, the computed robustness margin becomes tight. Let us mention that there exist many approaches to compute $\delta$-overapproximations of reachable sets for any desired accuracy $\delta$ for both linear \cite{le2010reachability} and nonlinear systems \cite{rungger2018accurate}. An illustration of the computation of the robustness margin is provided in Figure \ref{fig:robustness_margin}.

\subsection{Uniform robustness margins}

We have the following corollary of Theorem \ref{thrm:main} providing a uniform robustness margin $\varepsilon \geq 0$ for the system $\Sigma$ in (\ref{eqn:system}).

\begin{corollary}
    Consider the discrete-time control system $\Sigma$ in (\ref{eqn:system}) and its associated symbolic model $S_d(\Sigma)=(X_d, X_d^0, U_d, \Delta_d)$ constructed in Section \ref{Def:Sd}. Define $\varepsilon:=\min\limits_{q \in X_d,v \in U_d}\eta(q,v,A)$ with $A=\Delta_d(q,v)$. Consider the relation $\mathcal{R} \subseteq X\times X_d$ defined by 
    \begin{equation}
    \label{eqn:relation}
        (x, q) \in \mathcal{R} \text{ if and only if } x \in q.
        \end{equation}
    Then, the following holds:
    \begin{itemize}
        \item $\varepsilon>0$ and $\mathcal{R}$ is an alternating simulation relation from $S_d(\Sigma)$ to $S(\Sigma_{\mu})$ for any $\mu \geq 0$, such that $\varepsilon > \mu$;
        \item For $\delta \geq 0$, if $\overline{f}$ is a $\delta$-overapproximation of the map $f$ describing the dynamics of the system $\Sigma$, and the set $\overline{f}(q,v,D)$ is closed for all $(q,v) \in X_d \times U_d$, then for all $\overline{\mu}> \varepsilon+\delta$, $\mathcal{R}$ is not an alternating simulation relation from $S_d(\Sigma)$ to $S(\Sigma_{\bar{\mu}})$.
    \end{itemize}
\end{corollary}
\begin{proof}
We show each item separately. We have shown in Theorem \ref{thrm:main} that $\eta(q,v,A)>0$ for all $(q,v) \in X_d\times U_d$. Since the sets $X_d$ and $U_d$ are finite, one gets that $\varepsilon=\min\limits_{q \in X_d,v \in U_d}\eta(q,v,A)>0$.

To show the behavioral relationship between $S_d(\Sigma)$ and $S(\Sigma_{\mu})$, one can follow the same steps used in the proof of Theorem
\ref{thrm:main}.

To show that $\mathcal{R}$ in (\ref{eqn:relation}) is not an alternating simulation relation from  $S_d(\Sigma)$ to $S(\Sigma_{\overline{\mu}})$, we show that the second item in Definition \ref{Def:altsimu} does not hold.
Since $\varepsilon=\min\limits_{q \in X_d,v \in U_d}\eta(q,v,\Delta_d(q,v))$, we have the existence of $q_m \in X_d$ and $u_m \in U_d$ such that $\varepsilon=\eta(q_m,u_m,A)$ with $A = \Delta_d(q_m,u_m)$. By definition of the map $\eta$ in (\ref{Def:eta_a}) and since $\varepsilon=\eta(q_m,u_m,A)$, we have the existence of $z \in q_m$ such that for all $\varepsilon_0 > \varepsilon $
\begin{equation}
\{\cl(\overline{f}(z, u_m, D)) +  \mathcal{B}_{\varepsilon_0}(0) \} \cap \mathbb{R}^n \setminus Q^{-1}(\Delta_d(q_m, u_m)) \neq \emptyset.  
\end{equation}
In particular for $\varepsilon_0=\overline{\mu}-\delta> \varepsilon$, we have that: 
\begin{equation}
\label{eqn:relation4}
\{\cl(\overline{f}(z, u_m, D)) +  \mathcal{B}_{\overline{\mu}-\delta}(0) \} \cap \mathbb{R}^n \setminus Q^{-1}(\Delta_d(q_m, u_m)) \neq \emptyset  
\end{equation}
Moreover, since $\overline{f}$ is a $\delta-$overapproximation of $f$ and the overapproximation map $\overline{f}$ is closed, we have:
\begin{align*}
\cl(\overline{f}(z,u_m,D)) + \mathcal{B}_{\overline{\mu}-\delta}(0)  &= \overline{f}(z,u_m,D) + \mathcal{B}_{\overline{\mu}-\delta}(0) \\  &\subseteq f(z,u_m,D) + \mathcal{B}_{\overline{\mu}}(0)
\end{align*}
Combining the last inclusion with (\ref{eqn:relation4}) we deduce that $  f(z,u_m,D) +  \mathcal{B}_{\overline{\mu}}(0) \cap \mathbb{R}^n \setminus Q^{-1}(\Delta_d(q_m, u_m)) \neq \emptyset$. Hence, there exists $y \in \{f(z,u_m,D)+\mathcal{B}_{\overline{\mu}}(0)\}$ such that for all $q' \in \Delta_d(q_m,u_m)$, $y \notin q'$. Thus, $\mathcal{R}$ is not an alternating simulation relation from $S_d(\Sigma)$ to $S(\Sigma_{\bar{\mu}})$.

\end{proof}

\subsection{Controller synthesis for perturbed systems}

We now introduce the concept of controller for a transition system. Consider the system $S=\left(X, X^0, U, \Delta\right)$ and the controller $\mathcal{C}: X \rightrightarrows U$ such that for all $x \in X, \mathcal{C}(x) \subseteq U_S^a(x)$. Let $dom(\mathcal{C})$ be the domain of the controller defined by $dom(\mathcal{C})=\{x \in X \mid \mathcal{C}(x) \neq \emptyset\} \subseteq X$. We define a controlled transition system by a tuple $S^{\mathcal{C}} =\left(X_\mathcal{C}, X_\mathcal{C}^0, U_\mathcal{C}, \Delta_\mathcal{C}\right)$, where:
\begin{itemize}

    \item $X^\mathcal{C}=X \cap dom(\mathcal{C})$ is the set of states;
    \item $X^{\mathcal{C}, 0}=X^0 \cap dom(\mathcal{C})$ is the set of initial states;
    \item $U^\mathcal{C}=U$ is the set of inputs;
    \item   A transition relation: $x_\mathcal{C}^{\prime} \in \Delta_\mathcal{C}\left(x_\mathcal{C}, u_\mathcal{C}\right)$ if and only if $x_\mathcal{C}^{\prime} \in \Delta\left(x_\mathcal{C}, u_\mathcal{C}\right)$ and $u_\mathcal{C} \in$ $\mathcal{C}\left(x_\mathcal{C}\right)$.
\end{itemize}
Given a specification $\mathcal{H}_{Spec}\subseteq \mathcal{H}_{x}(S)$, $\mathcal{C}$ is said to be a controller for the system $S$ and specification $\mathcal{H}_{Spec}$ if $\mathcal{H}_x(S^\mathcal{C}) \subseteq \mathcal{H}_{Spec}$.

We have the following result, showing the refinement procedure for the controller from the disturbance-free symbolic abstraction to the perturbed original system.

\begin{proposition}
\label{prop:2}
Consider the discrete-time control system $\Sigma$ in (\ref{eqn:system}) and its associated symbolic model $S_d(\Sigma)=(X_d, X_d^0, U_d, \Delta_d)$ constructed in section \ref{Def:Sd}. Let \( \mathcal{C}_{d} \) be a controller for the system $S_{d}(\Sigma)$ and specification $B_{spec}$. Consider the map $\varepsilon:X \times U_d \rightarrow \mathbb{R}_{\geq 0}$, defined for $(x,u) \in X \times U_d$ as $\varepsilon(x,u)=\eta(Q(x),u,A)$, with $A=\Delta_d(Q(x),u)$. Consider a map $\mu:X \times U_d \rightarrow \mathbb{R}_{\geq 0}$ satisfying $\mu(x,u) < \varepsilon(x,u)$ for all $(x,u) \in X \times U_d$. We define the controller $\mathcal{C}:X\rightrightarrows U_d$ as follows, for $x \in X$, $\mathcal{C}(x) = \mathcal{C}_{d}(Q(x))$. Then \(\mathcal{C}\) is a controller for the perturbed system \(S(\Sigma_{\mu})\) and the specification $Q^{-1}(B_{spec})$.
\end{proposition}

\begin{proof}
Let \(\sigma = ((x_0,u_0), (x_1,u_1), \ldots) \in \mathcal{H}( S^\mathcal{C}(\Sigma_{\mu}))\) where \(x_0 \in \dom(\mathcal{C})\). Let us show the existence of a behavior \(\sigma_d = ((q_0,u_0), (q_1,u_1), \ldots) \in  \mathcal{H}( S^{\mathcal{C}_{d}}_{d}(\Sigma))\) with $q_i = Q(x_i)$ for all $i \in \N$.
To show this result, we need to show that for all \(i \in \N\), we have $u_i \in  \mathcal{C}_{d}(q_i)$ and $q_{i+1} \in \Delta_{d}(q_i, u_i)$.
Let \(i \in \N\), first we have \(u_i \in \mathcal{C}(x_i) = \mathcal{C}_{d}(Q(x_i)) = \mathcal{C}_{d}(q_i)\) therefore the first condition is satisfied.
For the second condition, we need to show that \(q_{i+1} \in \Delta_{d}(q_i, u_i)\) which is equivalent to proving that \(\{\cl(\overline{f}(q_i,u_i,D))\} \cap \cl(q_{i+1}) \neq \emptyset\).
Using the fact that \( \mu(x_i, u_i) < \varepsilon(x_i, u_i)\), one gets \( \cl(\overline{f}(x_i,u_i,D))+\mathcal{B}_{\mu(x_i,u_i)}(0) \subseteq \cl(\overline{f}(q_i,u_i,D))+\mathcal{B}_{\varepsilon(q_i,u_i)}(0) \subseteq \Int(Q^{-1}(\Delta_d(q_i, u_i))) \). Thus, for  \(x_{i+1} \in f(x_i,u_i,D)+\mathcal{B}_{\mu(x_i,u_i)}(0)\), it follows that \(x_{i+1} \in \Int(Q^{-1}(\Delta_d(q_i, u_i)))\). Therefore, \( Q(x_{i +1 }) = q_{i+1} \in \Delta_{d}(q_i, u_i)\), which concludes the proof of the two conditions.
Moreover, we know that $x_0 \in \dom(\mathcal{C})$ thus \(q_0=Q(x_0) \in \dom(\mathcal{C}_{d})\). Therefore, from the fact that $\mathcal{C}_{d}$ is a controller for the system $S_d(\Sigma)$ and specification $B_{spec}$ it follows $\sigma_{d,x}=(q_0,q_1,q_2,\ldots) \in B_{spec}$. Hence, we deduce that $\sigma_x=(x_0,x_1,x_2,\ldots) \in Q^{-1}(B_{spec})$. Hence, \(\mathcal{C}\) is a controller for the perturbed system \(S(\Sigma_{\mu})\) and the specification $Q^{-1}(B_{spec})$.
\end{proof}

\subsection{Controller determination to maximize the robustness margin}
\label{max_controller}

Given the discrete-time dynamical control system $\Sigma$ in (\ref{eqn:system}) and its associated symbolic model $S_d(\Sigma)=(X_d, X_d^0, U_d, \Delta_d)$ constructed in Section \ref{Def:Sd}. In the previous section, we have shown how to construct a controller $\mathcal{C}: X\rightrightarrows U_d$ for the perturbed system to satisfy a given specification. Since the controller $\mathcal{C}$ is nondeterministic, one can determinize it while further improving the obtained robustness margin. In this context, we introduce the deterministic controller $\overline{\mathcal{C}}:X \rightarrow U_d$ can be defined by:
\begin{equation}
\label{eqn:deter_cont}
    \overline{\mathcal{C}}(x)=\argmax_{u \in \mathcal{C}(x)}\eta(Q(x),u,\Delta_d(Q(x),u)).
    \end{equation}
Indeed, for a given state $x$, the controller chooses the control input $u \in \mathcal{C}(x)$ that maximizes the admissible robustness margin, that is $\eta(Q(x),u,\Delta_d(Q(x),u)) \geq \eta(Q(x),u',\Delta_d(Q(x),u'))$ for all $u' \in \mathcal{C}(x)$.

\section{Numerical example}
\label{sec:4}

In this section, we present numerical examples to illustrate the theoretical part of our work. We consider two examples, a first example of a double integrator with a safety specification and a mobile robot example with a Co-Safe LTL specification.

\subsection{Double integrator}

We consider a discrete-time version of a double integrator described by:
$$
\left\{\begin{array}{l}
x^1_{k+1} = x^1_k + \tau x^2_k + \frac{\tau^2}{2} u_k + \frac{\tau^2}{2} w^1_k, \\
x^2_{k+1} = x^2_k + \tau u_k + \tau w^2_k
\end{array}\right.
$$

where $x=(x^1,x^2) \in X =[0, 6]\times[-6,  6]$ represents the state of the system, consisting of the position and the velocity, $u \in [-1, 1]$ represents the control input, and  $w=(w^1,w^2)\in [-0.01, 0.01] \times  [-0.01, 0.01]$ represents the disturbance inputs affecting the system. For this example, the sampling period is chosen as $\tau= 0.5$ seconds.

We construct a symbolic model of the considered system following the approach described in Section \ref{Def:Sd}. First, we uniformly discretize the state space \(X \) into \( N_{x_1} = 80 \) subintervals along the first axis and \( N_{x_2} = 160 \) subintervals along the second axis. The control input is uniformly discretized into \( N_{u_1} = 5 \) values. We then compute the transition relation of the symbolic model by using reachability analysis. For the computation of the reachable sets, since we are considering a system with linear dynamics and interval-type partition, the exact reachable set can be determined by evaluating only the extreme points of the state and disturbance input (see \cite{althoff2021set} for example). Finally, the admissible non-uniform robustness margin $\varepsilon: X \times U_d \rightarrow \mathbb{R}_{\geq 0}$ can be computed according to equation (\ref{Def:eta_a}) and Theorem \ref{thrm:main}. One can see that the admissible robustness margin belongs to the set $[0.01125,0.03625]$, and the uniform robustness margin is equal to $0.01125$. Table \ref{tab:my_table} represents the admissible uniform robustness margin for different state and input discretization parameters, one can see that the finer is the chosen discretization, the smaller is the admissible uniform robustness margin, which reflects the trade-off between the accuracy and robustness of the computed symbolic abstraction.

\begin{figure}[htbp]
        \centering
        \includegraphics[scale=0.48]{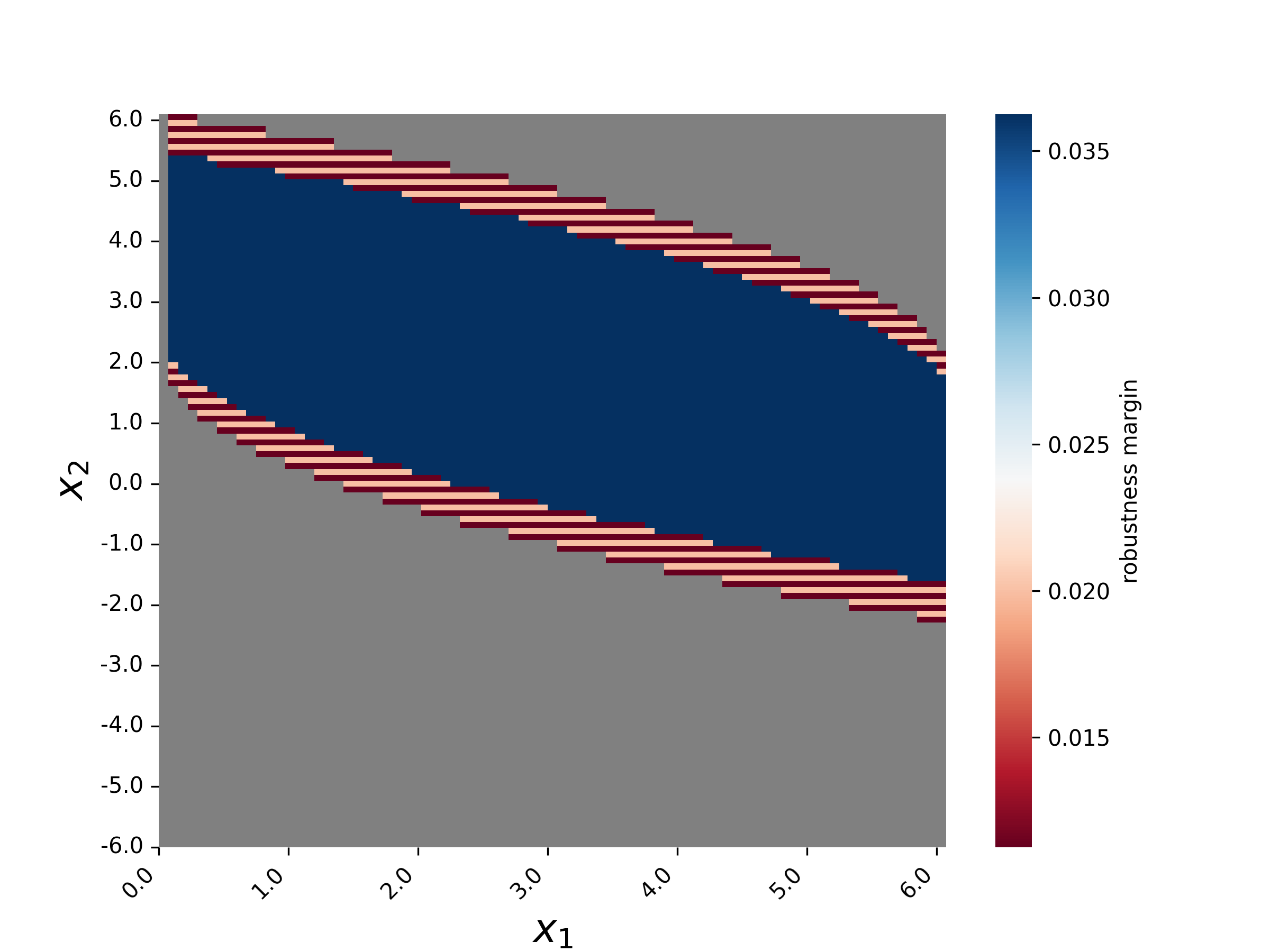}
        \caption{The maximal admissible robustness margin for each state.}
        \label{fig:robust_heatmap}
 \end{figure}

\begin{figure}[htbp]
        \centering
        \includegraphics[scale=0.48]{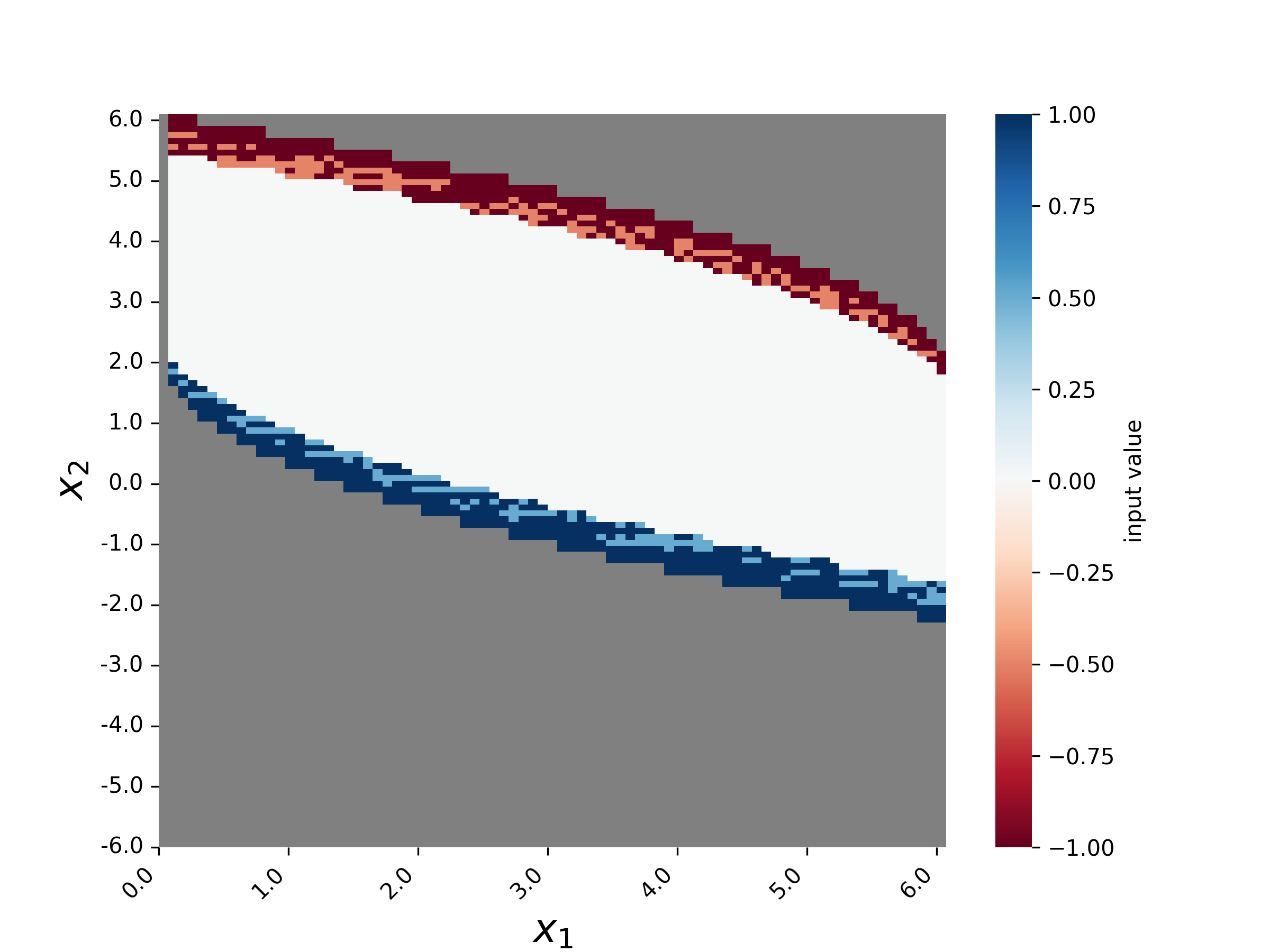}
        \caption{The input value maximizing the robustness margin for each state.}
        \label{fig:input_heatmap}
\end{figure}

We then use the fixed-point algorithm \cite{tabuada2009verification} to construct the maximal safety controller $\mathcal{C}$ for the symbolic model $S_d(\Sigma)$ and safe set \( [0, 6] \times [-6, 6] \) as in \ref{max_controller}. In Figure \ref{fig:robust_heatmap}, we show the maximal admissible robustness margin for each state $x \in X$ by the map defined by $\eta(x)=\max\limits_{u \in \mathcal{C}(x)} \varepsilon(x,u)$. Indeed, we are only limiting the computation of the robustness margin to the states belonging to the domain of the controller. Figure \ref{fig:input_heatmap} represents the deterministic controller $\overline{\mathcal{C}}$ defined in (\ref{eqn:deter_cont}) representing for each state, the input value that maximizes the robustness margin, as described in Section \ref{max_controller}.

\begin{figure}[htbp]
    \centering
    \includegraphics[scale=0.48]{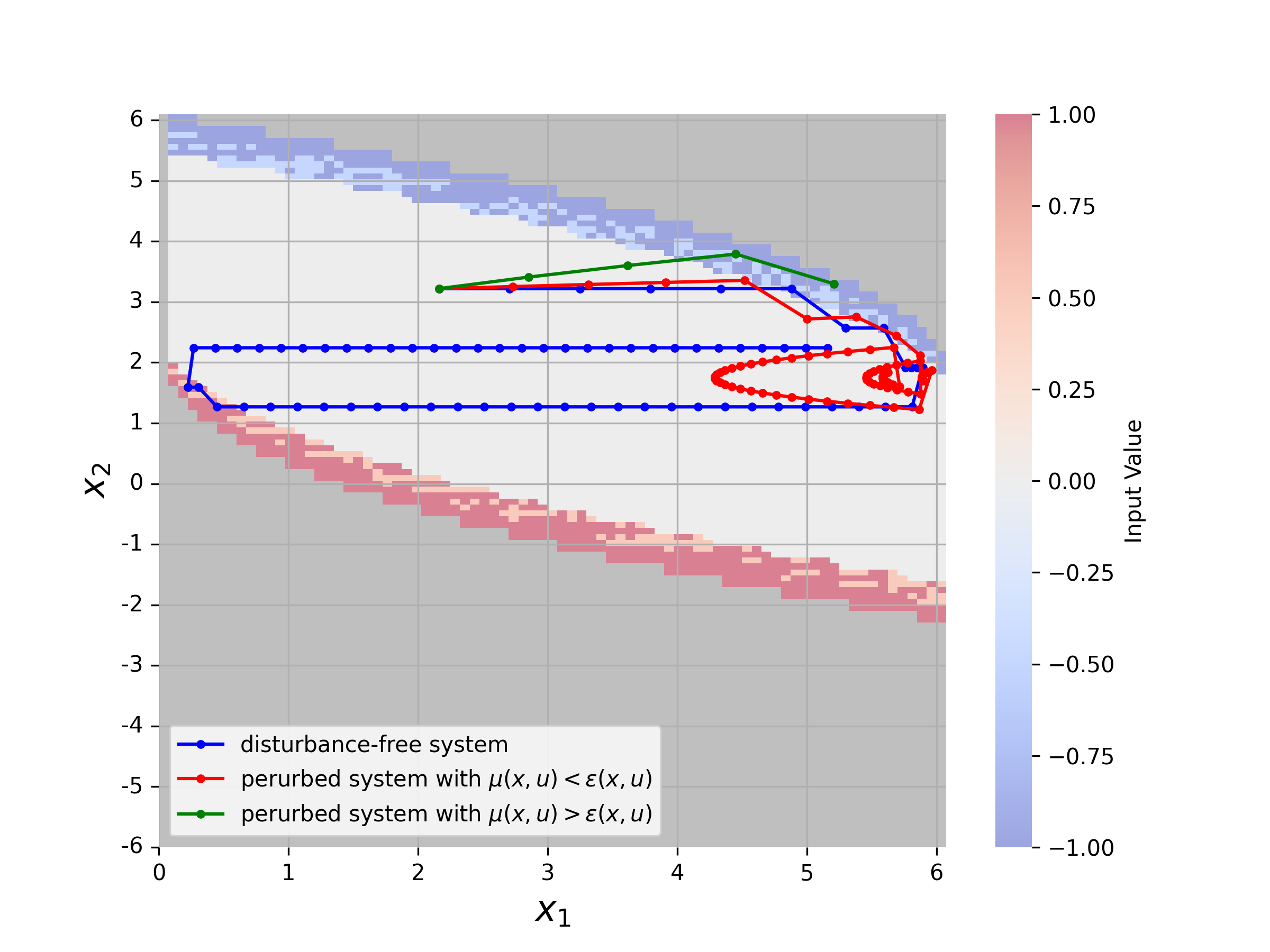}
    \caption{Three trajectories of the dynamics starting from the initial state $x_0=(2.2, 3.2)$ for a $100$ time step, under different perturbations.}
    \label{fig:trajectory_heatmap}
\end{figure}

Figure \ref{fig:trajectory_heatmap} represents the trajectories of the closed-loop system starting from the initial condition $x_0=(2.2, 3.2)$. For the simulations, we use the deterministic controller $\overline{\mathcal{C}}$ that maximizes the robustness margin defined in (\ref{eqn:deter_cont}). The blue trajectory represents the evolution of the closed-loop disturbance-free system. The red trajectory represents the evolution of the closed-loop disturbed system, where the considered additive disturbance $\mu: X \times U_d \rightarrow \mathbb{R}_{\geq 0}$ satisfy (\ref{eqn:robu1}), one can see that the trajectory of the perturbed system remains in the safe set, which is consistent with the second item of Theorem \ref{thrm:main} and Proposition \ref{prop:2}. Finally, The green trajectory represents the evolution of the closed-loop disturbed system starting from $x_0$, where the considered additive disturbance $\mu: X \times U_d \rightarrow \mathbb{R}_{\geq 0}$ satisfy condition (\ref{eqn:robu2}), where the parameter of the overapproximation of the reachable set is given by $\delta=0$, since the reachable set can be computed exactly for this example. One can see that after five steps, the trajectory leaves the domain of the controller.

\begin{table}[h!]
\centering
\begin{tabular}{|c|c|c|c|c|}
\hline
\diagbox[width=1cm, height=1cm]{$N_u$}{$N_x$}& 1600 & 3200 & 6400 & 12800 \\
\hline
3 & 0.0237  & 0.0237  & 0.0237 & 0.0112\\
\hline
5 &  0.0237 & 0.01125&  0.01125& 0.01125\\
\hline
10 & 0.012638  & 0.0043  & 0.0043 & 0.00291\\
\hline
\end{tabular}
\caption{Uniform robustness margin for different space discretization:
$N_x = N_{x_1}\times N_{x_2}$ (total number of discrete states) and $N_u$ (number of symbolic
inputs).}
\label{tab:my_table}
\end{table}


\subsection{Mobile robot}

In this second example, we consider a unicycle-type robot. Its state is characterized by its position $\left(x^1, x^2\right) \in\left[0, 10\right] \times\left[0, 10\right]$ and its orientation \(x^3 \in [-\pi, \pi]\). The discrete-time dynamics of the system are defined by the following equations:

$$
\left\{\begin{array}{l}
x^1_{k+1}=x^1_k+\tau\left(u^1_k \cos \left(x^3_k\right)+w^1_k\right) \\
x^2_{k+1}=x^2_k+\tau\left(u^1_k \sin \left(x^3_k\right)+w^2_k\right) \\
x^3_{k+1}=x^3_k+\tau\left(u^2_k+w^3_k\right)(\bmod 2 \pi)
\end{array}\right.
$$
where  $(u^1,u^2) \in\left[0.25, 1\right]\times [-1, 1]$ are the control inputs of the system representing the robot's linear and angular velocities, $w^i(t) \in\left[-0.05, 0.05\right]$ for $ i=1,2,3$  are disturbances, and \( \tau =1\) is the sampling period. 

\begin{figure}[htbp]
    \centering
    \includegraphics[width=0.5\textwidth]{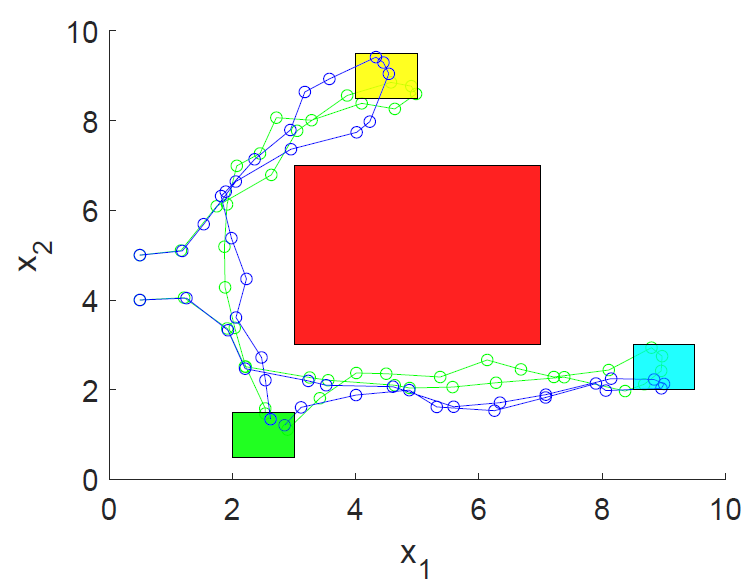}
    \caption{Simulation of three trajectories of robot dynamic with and without perturbation starting from two initial states $x_0=(0.5, 4)$ and $x_1 = (0.5, 5)$. }
    \label{fig:traject_without_p}
\end{figure}

To describe the control objective, we first define the four disjoint regions \( R_1, R_2, R_3, R_4 \) within the domain \( \mathbb{X} = [0, 10] \times [0, 10] \times [-\pi, \pi] \), which are illustrated in Figure \ref{fig:traject_without_p} and whose numerical values are provided in Table \ref{fig:regions}. The considered specification can be described as follows: reach either \( R_1 \) or \( R_2 \), but only one of them, then reach \( R_3 \), while avoiding \( R_4 \) throughout the entire path.

We construct a symbolic model of the considered system following the approach described in Section \ref{Def:Sd}. We consider a Cartesian partition of the state space \( \mathbb{X} \), uniformly dividing it into \( N_{x_1} = 100 \), \( N_{x_2} = 100 \), and \( N_{x_3} = 30 \) subintervals along each axis, respectively. The control inputs are uniformly discretized into \( N_{u_1} = 3 \) and \( N_{u_2} = 5 \) values. 

We then compute the transition relation of the symbolic model by using reachability analysis. For the computation of the reachable sets, we use the growth bounds-based approach proposed in \cite{ReissigWeberRungger17}. Finally, the admissible non-uniform robustness margin $\varepsilon: X \times U_d \rightarrow \mathbb{R}_{\geq 0}$ can be computed according to equation (\ref{Def:eta_a}) and Theorem \ref{thrm:main}. The value of the admissible robustness margin belongs to the ranges $[0.0021, 0.725]$, $[0.0011, 0.05]$, and $[0.0311, 0.1122]$ for each respective dimension. The mean robustness margin in each direction is equal to $[0.0025, 0.0017, 0.0311]$. 

Since the considered specification belongs to the class of Co-Safe LTL specification~\cite{girard2024approches}, we used the algorithmic tools presented in \cite{belta2017formal} for controller synthesis.

In Figure \ref{fig:traject_without_p}, we show blue trajectories obtained for two distinct initial conditions of the closed-loop disturbance-free system from $x_0=(0.5, 4)$ and $x_0=(0.5, 5)$. We can observe that the specification is satisfied: the first trajectory reaches \( R_1 \) and then reaches \( R_3 \) without passing through \( R_2 \); the second trajectory reaches \( R_2 \) and then reaches \( R_3 \) without passing through \( R_1 \). In both cases, \( R_4 \) is avoided throughout the trajectory. In green, we show trajectories starting from $x_0 = (0.5, 4)$ and $x_0=(0.5, 5)$, that are subject to an additive disturbance $\mu: X \times U_d \rightarrow \mathbb{R}_{\geq 0}$ satisfying condition (\ref{eqn:robu1}). One can see that the control objective is achieved, which is consistent with the second item of Theorem \ref{thrm:main} and Proposition \ref{prop:2}.

\begin{table}[h!]
\centering
\renewcommand{\arraystretch}{1.5} 
\begin{tabular}{|c|c|}
\hline
 $R_1$ & $[4, 5] \times [8.5, 9.5] \times [-\pi, \pi] $ \\
\hline
$R_2$ & $[8.5, 9.5] \times [2, 3] \times [-\pi,  \pi] $ \\
\hline
$R_3$ &  $[2, 3] \times [0.5, 1.5] \times [-\pi, \pi] $ \\
\hline
$R_4$ & $[3, 7] \times [3, 7] \times [-\pi, \pi] $  \\
\hline
\end{tabular}
\caption{Numerical values of the regions involved in the specification $\mathcal{S}$.}
\label{fig:regions}
\end{table}

\section{Conclusion}

We have introduced a new concept of maximal robustness margin for discrete-time dynamical systems. This quantity represents the largest disturbance that can be introduced into a dynamical system while preserving the alternating simulation relation with the disturbance-free abstract model. Our work distinguishes between non-uniform robustness margins (state and input dependent) and uniform robustness margins. We then show how to exploit robustness margin for the sake of controller synthesis. Simulations are performed on two numerical examples to demonstrate the effectiveness and performance of the proposed approach, though the results show these margins can be conservative for systems with small state/input spaces and disturbance sets.
The proposed robustness margins in this paper make it possible to preserve the behavioral relationship between the original system and its symbolic model, making them applicable for controller synthesis under any specification. In future work, we aim to provide refined robustness margins tailored to specific specifications. By computing robustness margins for the predecessor operator, the margins proposed here can be improved to address particular specifications, such as safety, reachability, persistence, or recurrence, while potentially reducing their conservatism.

\bibliographystyle{cas-model2-names}

\bibliography{cas-refs}

\end{document}